\documentclass[conference]{IEEEtran}

\ifCLASSINFOpdf
\else
\fi

\usepackage{bm}
\usepackage{floatflt}

\usepackage{color}
\usepackage{cite}

\usepackage{graphicx}
\usepackage{amsmath}
\usepackage{amsthm}
\usepackage{amssymb}

\usepackage{verbatim}

\usepackage{psfrag,array}

\usepackage{subfigure}

\usepackage{stfloats}

\usepackage{todonotes}

\usepackage{amsmath}
\usepackage{epstopdf}
\usepackage[ruled,vlined]{algorithm2e}
\usepackage{booktabs}

\newcommand{\p}[1]{\mathop{\mbox{\it p} } }

\newcommand{\be}{\begin{equation}}
\newcommand{\ee}{\end{equation}}
\newcommand{\ba}{\begin{array}}
\newcommand{\ea}{\end{array}}
\newcommand{\bea}{\begin{eqnarray}}
\newcommand{\eea}{\end{eqnarray}}
\newcommand{\bean}{\begin{eqnarray*}}
\newcommand{\eean}{\end{eqnarray*}}

\definecolor{white}{rgb}{1,1,1}

\newtheorem*{theorem}{Theorem}

\usepackage{xcolor}
\newenvironment{Proof}{{\noindent\it Proof:~}}{\hfill $\square$\par}
\begin{document}
\title{Minimum Eigenvalue Based Covariance Matrix Estimation with Limited Samples}

\author{\IEEEauthorblockN{Jing Qian, Juening Jin and Hao Wang}
\IEEEauthorblockA{Huawei Technologies Co., Ltd., Beijing, China\\
Email: qianjing3@huawei.com, jinjuening@hisilicon.com, hunter.wanghao@huawei.com}
}

%
\maketitle


\begin{abstract}
In this paper, we consider the interference rejection combining (IRC) receiver, which improves the cell-edge user throughput via suppressing inter-cell interference and requires estimating the covariance matrix including the inter-cell interference with high accuracy. In order to solve the problem of sample covariance matrix estimation with limited samples, a regularization parameter optimization based on the minimum eigenvalue criterion is developed. It is different from traditional methods that aim at minimizing the mean squared error, but goes straight at the objective of optimizing the final performance of the IRC receiver. A lower bound of the minimum eigenvalue that is easier to calculate is also derived. Simulation results demonstrate that the proposed approach is effective and can approach the performance of the oracle estimator in terms of the mutual information metric.
\end{abstract}


%
\IEEEpeerreviewmaketitle

\section{Introduction}
It is well known that high-capacity cellular systems are interference limited. In addition to the cell-edges, the interference scenario is becoming even more complex with the introduction of heterogeneous systems involving macro base stations and pico base stations, device-to-device communications and cognitive radios. Algorithms that deal with interference have been extensively considered and are expected to play a vital role in the cellular systems. One of the important techniques of such kind is interference rejection combining (IRC) which essentially performs effective inter-cell interference suppression \cite{ref:irc1,ref:irc2}. Such a receiver employs multiple antennas to spatially suppress the interfering signal and is considered to be highly beneficial for the existing and future 5th generation (5G) systems.

In the IRC receiver, the correlation of the interference among the receive antennas is utilized to perform interference rejection and combining, and it requires the knowledge of interference signals, i.e., the covariance matrix including the interference signals and noise. As a consequence, the IRC receiver is sensitive to not only the channel estimation error but also the covariance matrix estimation error.

Based on the maximum likelihood (ML) estimation principle and mutually independence between received signals, the unbiased sample covariance matrix (SCM) turns out to be the most used estimator in lots of applied problems \cite{ref:irc7,ref:irc8}. However, it is known that SCM becomes less and less accurate as the number of samples $M$ decreases compared to the dimensionality of the sample space $N$. On one hand, it performs poorly for $M<N$ since it is not possible to invert the sample covariance matrix. On the other hand when $M\geq N$, the covariance matrix can be inverted, but inverting it leads to an even bigger estimation error, hence, it is ill conditioned.


Better estimators can be developed by using regularization, where the key idea is to shift or shrink the estimator toward a predetermined target or model. This can significantly decrease the
variance of the estimator and improve the overall performance by reducing its mean squared error (MSE). \cite{ref:irc7} investigated an expected likelihood based regularization approach to force the estimated covariance matrix to have the same statistical quality as the unknown actual covariance matrix. An empirical Bayes approach was proposed in \cite{ref:irc9} that aims at the optimizing covariance-matrix-specific metrics.

However, as IRC performs as a receiver, the MSE of the covariance matrix may not be directly related to the final performance, i.e., block error rate. Therefore, we propose to optimize the regularization in order to maximize the signal to interference plus noise ratio (SINR), or equivalently, to minimize the difference between the true SINR and the regularized SINR. Furthermore, we derive that the regularization can be optimized via making the minimum eigenvalue of the regularized SCM as precise as possible.

In general, regularization combines an unstructured estimate with a predefined model or target, which can be decided based on a prior knowledge, or on the properties that we want to enforce. In this paper, we also demonstrate that the target of regularization from the perspective of eigenvalue optimization.

Automatic setting of the regularization parameter requires to adopt a suitable criterion, which may depend on the application; moreover, the optimal value typically depends on the true but unknown covariance matrix \cite{ref:irc3}. Implementable methods that have been proposed often rely on approximations or, alternatively, involve computational-intensive procedures \cite{ref:irc4, ref:irc5, ref:irc6}. In this paper, we derive a lower bound of the minimum eigenvalue that is easier to calculate to approximate itself.

Finally, we report on the performance comparison between the proposed estimator and two competitors: the ideal and non-computable oracle estimator and the unregularized ML estimator. Results show that a significant gain can be obtained in terms of the mutual information metric, which is a representative for the BLER performance.

The remainder of this paper is structured as follows. Sec. II describes the system model. In Sec. III, the proposed regularization enhancements are elaborated. The performance of optimized covariance matrix is evaluated in Sec. IV. Sec. V concludes the paper.

\section{Preliminaries}
\subsection{System Model}
\begin{figure}[t]
\centering
\includegraphics[width=3in]{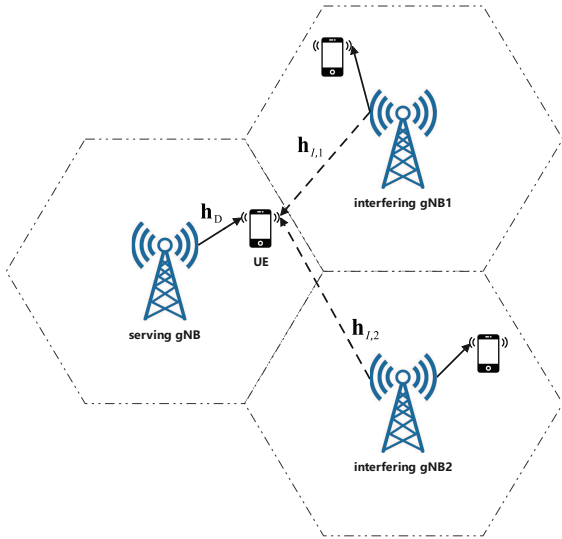}
\caption{Cell configuration, where $N_D=1,N_I=2$.}
\label{fig:cell}
\vspace{-4mm}
\end{figure}

Considering an Bit Interleaved Coded Modulation (BICM) and MIMO system with flat fading channels and $N_R$ receiving antennas. Let us assume that the subscripts $D$ and $I$ denote the desired signal and the interfering signal, respectively, and $\mathbf{s}_D\in\mathbb{C}^{N_D\times1}$ and $\mathbf{s}_I\in\mathbb{C}^{N_I\times1}$ are denoted as the desired transmitted signal and the interfering transmitted signal, respectively. For the desired transmitted signal, each $\log_2 Q$ bits are mapped into a quadrature amplitude modulation (QAM) symbol from a constellation $\mathcal{Q} = \{X_i,i=1,\cdots,Q\}$. After transmission over the fading channels, the frequency domain received signal vector at the $N_R$ receiving antennas can be expressed as
\begin{align}
\mathbf{y}=\mathbf{H}_D\mathbf{s}_D+\mathbf{H}_I\mathbf{s}_I+\mathbf{n}\triangleq\mathbf{H}_D\mathbf{s}_D+\mathbf{u},
\end{align}
where $\mathbf{H}_D\in \mathbb{C}^{N_R\times N_D}$ and $\mathbf{H}_I\in \mathbb{C}^{N_R\times N_I}$ are the equivalent channel matrices which include the precoding matrices, and $\mathbf{n}\in \mathbb{C}^{N_R\times1}\sim\mathcal{CN}(\mathbf{0},\mathbf{I})$ is the additive white Gaussian noise (AWGN) vector with unit power. The scenario shown in Fig. \ref{fig:cell} is taken as an example, where $N_I=2$.

Since the interference and noise are independent, the interference plus noise covariance matrix for $\mathbf{u}$ is given by
\begin{align}
\mathbf{R}=\mathbb{E}\{\mathbf{u}\mathbf{u}^\mathrm{H}\}=\mathbb{E}\{\mathbf{H}_I\mathbf{H}_I^\mathrm{H}\}+\mathbf{I},
\end{align}

The interference plus noise covariance matrix can be estimated by the sample covariance matrix of a set of received signals \cite{ref:irc10}, $\mathbf{u}(m)$, where $\mathbf{u}(m)\sim\mathcal{CN}(0, \hat{\mathbf{R}}_{\mathrm{ML}})$ and $m\in{1,2,\cdots,M}$, where $\hat{\mathbf{R}}_{\mathrm{ML}}$ represents the SCM. Based on the maximum likelihood estimation principle and mutually independence between received signals, the SCM can be expressed as
\begin{align}\label{eq:SCM}
\hat{\mathbf{R}}_{\mathrm{ML}}=\frac{1}{M}\sum_{m=1}^M\mathbf{u}(m)\mathbf{u}(m)^\mathrm{H}.
\end{align}
where $m$ is the sample index and $M$ is the observation length in subcarrier samples.

The estimator in (\ref{eq:SCM}) is a consistent one, in the sense that it converges to the true covariance matrix as $M\rightarrow \infty$. Its advantages are ease of computation and the property of being unbiased (i.e., its expected value is equal to the true covariance matrix). Its main disadvantage is the fact that it contains a lot of estimation error when the number of data points is of comparable or even smaller order than the dimension.

This problem is solved by coupling the pooled estimator with a spherical target matrix, that is, by regularization toward a scaled identity matrix via
\begin{align}\label{eq:regular}
\hat{\mathbf{R}}_{\mathrm{SE}}(\rho)=(1-\rho)\hat{\mathbf{R}}_{\mathrm{ML}}+\rho\frac{\mathrm{tr}(\hat{\mathbf{R}}_{\mathrm{ML}})}{N_R}\mathbf{I},~\rho\in [0,1].
\end{align}

%
%
%
%
%
\subsection{Eigenvalues of Sample Covariance Matrix}
As the SCM for $\mathbf{u}$ is an unbiased estimator, i.e., $\mathbb{E}\{\hat{\mathbf{R}}_{\mathrm{ML}}\}=\mathbf{R}$, it is asymptotically optimal when the number of samples $M$ is sufficiently large. However, if available samples are limited, the eigenvalues of $\hat{\mathbf{R}}_{\mathrm{ML}}$ may be far from the true values. To illustrate this point, we consider the maximum and minimum eigenvalues, respectively denoted as $\kappa(\cdot)$ and $\lambda(\cdot)$, as examples.

Since $\kappa(\hat{\mathbf{R}}_{\mathrm{ML}})$ is convex with respect to $\hat{\mathbf{R}}_{\mathrm{ML}}$ and $\lambda(\hat{\mathbf{R}}_{\mathrm{ML}})$ is concave with respect to $\hat{\mathbf{R}}_{\mathrm{ML}}$, according to the Jensen's inequality, we have
\begin{align}
&\mathbb{E}\{\kappa(\hat{\mathbf{R}}_{\mathrm{ML}})\}\geq \kappa(\mathbb{E}\{\hat{\mathbf{R}}_{\mathrm{ML}}\})= \kappa(\mathbf{R}),\\
&\mathbb{E}\{\lambda(\hat{\mathbf{R}}_{\mathrm{ML}})\}\leq \lambda(\mathbb{E}\{\hat{\mathbf{R}}_{\mathrm{ML}}\})= \lambda(\mathbf{R}).
\end{align}

As can be seen, it is more likely that the maximum eigenvalue of the ML estimator $\hat{\mathbf{R}}_{\mathrm{ML}}$ is larger than that of $\mathbf{R}$, and the minimum eigenvalue of $\hat{\mathbf{R}}_{\mathrm{ML}}$ is smaller than that of $\mathbf{R}$.

\begin{figure}[t]
\centering
\includegraphics[width=3in]{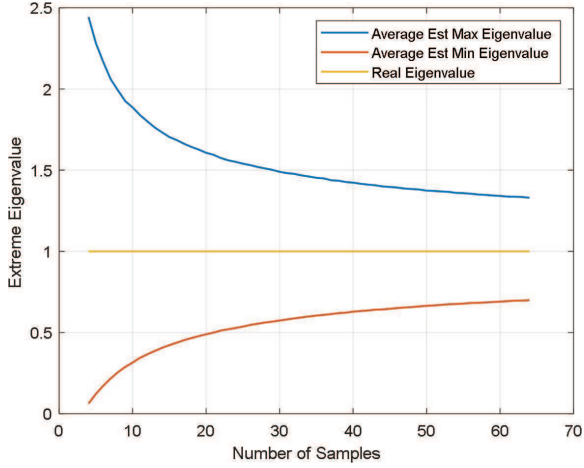}
\caption{Estimated eigenvalues of $\mathbf{R}=\mathbf{I}_{4\times4}$ with respect to the number of samples .}
\label{fig:eigen}
\vspace{-4mm}
\end{figure}

For example, we set $\mathbf{R}=\mathbf{I}_{4\times4}$, and the mean maximum eigenvalue and mean minimum eigenvalue of the ML covariance matrix estimator with different number of samples are illustrated in Fig. \ref{fig:eigen}. When the number of samples is twice the dimension of the vector, i.e., $M=8$, the estimated maximum eigenvalue is double the true value and the estimated maximum eigenvalue is a quarter the true value.

Thus, we have to revise the eigenvalues of $\hat{\mathbf{R}}_{\mathrm{ML}}$ in a way of moving towards the mean of eigenvalues, i.e., $\mathrm{tr}(\hat{\mathbf{R}}_{\mathrm{ML}})/N_R$, and the easiest method to achieve this goal is given by (\ref{eq:regular}).

The success of regularization depends on proper selection of the tuning parameter $\rho$.
\section{Minimum Eigenvalue Based Regularization}

In this section, we focus on the optimization problem of the regularization parameter, and propose a method which aims at maximizing the mutual information.

\subsection{Criterion of Minimum Eigenvalue}
Take the scenario with a single stream transmission of the desired signal as an example
\begin{align}
\mathbf{y}=\mathbf{h}_Ds_D+\mathbf{u}.
\end{align}

The IRC receiver utilizes the knowledge of the covariance matrix of the total interference plus noise and prewhitens the received signals via performing a Cholesky decomposition of the covariance matrix
\begin{align}
\hat{\mathbf{R}}_{\mathrm{SE}}(\rho)=\mathbf{L}\mathbf{L}^\mathrm{H}.
\end{align}

Then a pre-noise whitening filter is applied on the original received signal
\begin{align}
\mathbf{z}=\mathbf{g}s_D+\mathbf{v},
\end{align}
where $\mathbf{z}\triangleq\mathbf{L}^{-1}\mathbf{y}$, $\mathbf{g}\triangleq\mathbf{L}^{-1}\mathbf{h}_D$, $\mathbf{v}\triangleq\mathbf{L}^{-1}\mathbf{u}$ and $\mathbf{v}\sim\mathcal{CN}(\mathbf{0},\mathbf{I})$.

\begin{align}
\hat{s}_D=\frac{\mathbf{g}^\mathrm{H}\mathbf{z}}{\mathbf{g}^\mathrm{H}\mathbf{g}+1}
=\frac{\mathbf{h}_D^\mathrm{H}[\hat{\mathbf{R}}_{\mathrm{SE}}(\rho)]^{-1}\mathbf{y}}{\mathbf{h}_D^\mathrm{H}[\hat{\mathbf{R}}_{\mathrm{SE}}(\rho)]^{-1}\mathbf{h}_D+1}
\triangleq s_D+w,
\end{align}
where the variance of the equivalent interference and noise $u$ is given by
\begin{align}\label{eq:variance}
\mathbb{E}\{|w|^2\}=\frac{\mathbf{h}_D^\mathrm{H}[\hat{\mathbf{R}}_{\mathrm{SE}}(\rho)]^{-1}\mathbf{R}[\hat{\mathbf{R}}_{\mathrm{SE}}(\rho)]^{-1}\mathbf{h}_D}
{\left|\mathbf{h}_D^\mathrm{H}[\hat{\mathbf{R}}_{\mathrm{SE}}(\rho)]^{-1}\mathbf{h}_D+1\right|^2},
\end{align}

As the true covariance matrix $\mathbf{R}$ is unknown, (\ref{eq:variance}) is approximated as
\begin{align}
\mathbb{E}\{|w|^2\}=\frac{1}{\mathbf{h}_D^\mathrm{H}[\hat{\mathbf{R}}_{\mathrm{SE}}(\rho)]^{-1}\mathbf{h}_D},
\end{align}
then the actual SINR is given by
\begin{align}
\gamma=\mathbb{E}_{\mathbf{h}_D}\left\{\mathbf{h}_D^\mathrm{H}[\mathbf{R}_{\mathrm{SE}}(\rho)]^{-1}\mathbf{h}_D\right\}\approx\mathrm{tr}(\hat{[\mathbf{R}}_{\mathrm{SE}}(\rho)]^{-1}),
\end{align}

The optimal SINR can be expressed as
\begin{align}
\gamma^*=\mathbb{E}_{\mathbf{h}_D}\left\{\mathbf{h}_D^\mathrm{H}\mathbf{R}^{-1}\mathbf{h}_D\right\}\approx\mathrm{tr}(\mathbf{R}^{-1}).
\end{align}

In order to maximize the actual SINR, equivalently, minimize the difference between the actual and optimal SINR, which equals
\begin{align}
\rho^*=\arg\min_{\rho\in[0,1]}\mathbb{E}[\mathrm{tr}(\mathbf{R}^{-1})-\mathrm{tr}(\hat{[\mathbf{R}}_{\mathrm{SE}}(\rho)]^{-1})]^2,
\end{align}

As the matrix inversion is required after calculating the covariance matrix, and the maximum eigenvalue of $[\mathbf{R}_{\mathrm{SE}}(\rho)]^{-1}$ is $\frac{1}{\lambda(\mathbf{R}_{\mathrm{SE}}(\rho))}$, which means the minimum eigenvalue of $\mathbf{R}_{\mathrm{SE}}(\rho)$ has the greatest influence on the performance of the IRC receiver.

Thus, we propose the design of $\rho$ at the purpose of making the estimation of the minimum eigenvalue as precise as possible, which is given by
\begin{align}\label{eq:rho}
\rho^*=\arg\max_{\rho\in[0,1]} \mathbb{E}_{\hat{\mathbf{R}}_{\mathrm{SE}}}[\lambda(\mathbf{R})-\lambda(\hat{\mathbf{R}}_{\mathrm{SE}}(\rho))]^2.
\end{align}
where $\lambda(\cdot)$ denotes the minimum eigenvalue of the matrix.

\subsection{Derivation of Regularization Parameter}
According to (\ref{eq:regular}), the minimum eigenvalue of $\hat{\mathbf{R}}_{\mathrm{SE}}(\rho)$ can be expressed as
\begin{align}
\lambda(\hat{\mathbf{R}}_{\mathrm{SE}}(\rho))=(1-\rho)\lambda(\hat{\mathbf{R}}_{\mathrm{ML}})+\rho\frac{\mathrm{tr}(\hat{\mathbf{R}}_{\mathrm{ML}})}{N_R},
\end{align}
As the objective function in (\ref{eq:rho}) is a convex quadratic function with respect to $\rho$, we can obtain the closed form of the optimal solution $\rho^*$ as follows
\begin{align}
\!\!\!\rho^*=\frac{\mathbb{E}_{\hat{\mathbf{R}}_{\mathrm{ML}}}\left[\lambda(\mathbf{R})-\lambda(\hat{\mathbf{R}}_{\mathrm{ML}})\right]\!\!\left[\frac{\mathrm{tr}(\hat{\mathbf{R}}_{\mathrm{ML}})}{N_R}-\lambda(\hat{\mathbf{R}}_{\mathrm{ML}})\right]}
{\mathbb{E}_{\hat{\mathbf{R}}_{\mathrm{ML}}}\left[\frac{\mathrm{tr}(\hat{\mathbf{R}}_{\mathrm{ML}})}{N_R}-\lambda(\hat{\mathbf{R}}_{\mathrm{ML}})\right]^2}.
\end{align}

Since the minimum eigenvalue $\lambda(\hat{\mathbf{R}}_{\mathrm{ML}})$ is generally much smaller than the mean of eigenvalues $\frac{\mathrm{tr}(\hat{\mathbf{R}}_{\mathrm{ML}})}{N_R}$, $\rho^*$ can be approximated as
\begin{align}\label{eq:optrho}
\rho^*=\frac{\left[\lambda(\mathbf{R})-\mathbb{E}_{\hat{\mathbf{R}}_{\mathrm{ML}}}\lambda(\hat{\mathbf{R}}_{\mathrm{ML}})\right]\mathbb{E}_{\hat{\mathbf{R}}_{\mathrm{ML}}}\frac{\mathrm{tr}(\hat{\mathbf{R}}_{\mathrm{ML}})}{N_R}}
{\mathbb{E}_{\hat{\mathbf{R}}_{\mathrm{ML}}}\left[\frac{\mathrm{tr}(\hat{\mathbf{R}}_{\mathrm{ML}})}{N_R}\right]^2}.
\end{align}

Note that $\mathbb{E}_{\hat{\mathbf{R}}_{\mathrm{ML}}}\lambda(\hat{\mathbf{R}}_{\mathrm{ML}})$ can be approximated as a scaled version of $\lambda(\hat{\mathbf{R}})$, thus (\ref{eq:optrho}) can be further approximated as
\begin{align}
\rho^*=\frac{\beta_M\lambda(\mathbf{R})\mathbb{E}_{\hat{\mathbf{R}}_{\mathrm{ML}}}\frac{\mathrm{tr}(\hat{\mathbf{R}}_{\mathrm{ML}})}{N_R}}
{\mathbb{E}_{\hat{\mathbf{R}}_{\mathrm{ML}}}\left(\frac{\mathrm{tr}(\hat{\mathbf{R}}_{\mathrm{ML}})}{N_R}\right)^2}.
\end{align}

According to the Jensen's inequality,
\begin{align}
\!\!\!\mathbb{E}_{\hat{\mathbf{R}}_{\mathrm{ML}}}\!\!\left(\!\frac{\mathrm{tr}(\hat{\mathbf{R}}_{\mathrm{ML}})}{N_R}\!\right)^2
\!\!\geq\!\left(\!\mathbb{E}_{\hat{\mathbf{R}}_{\mathrm{ML}}}\!\!\frac{\mathrm{tr}(\hat{\mathbf{R}}_{\mathrm{ML}})}{N_R}\!\right)^2
\!\!=\!\left(\!\frac{\mathrm{tr}(\mathbf{R})}{N_R}\!\right)^2\!,\!\!
\end{align}
we can derive the approximated expression for $\rho$ as follows
\begin{align}
\rho^*\approx\beta_M\frac{\lambda(\mathbf{R})}{\frac{\mathrm{tr}(\mathbf{R})}{N_R}}.
\end{align}

It is seen that $\mathbf{R}$ can not be obtained, but $\lambda(\mathbf{R})$ and $\frac{\mathrm{tr}(\mathbf{R})}{N_R}$ can be approximated with the SCMs with large and small sample intervals, respectively.
\begin{align}\label{eq:rho*}
\rho^*\approx\beta_M\frac{\lambda(\hat{\mathbf{R}}_{\mathrm{ML},1})}{\frac{\mathrm{tr}(\hat{\mathbf{R}}_{\mathrm{ML},2})}{N_R}}.
\end{align}

\subsection{Simplified Calculation of Minimum Eigenvalue}
Since the computation of the minimum eigenvalue is complex, several lower bounds of the minimal eigenvalue of a class of Hermitian matrices are investigated \cite{ref:irc11,ref:irc12}. In this section, a lower bound of the minimum eigenvalue that is easier to calculate is derived to approximate the minimum eigenvalue itself.
\begin{theorem}
For any positive semidefinite matrix $\mathbf{A}\in\mathbb{C}^{n\times n}$, we have
\begin{equation}
\begin{aligned}
&\kappa(\mathbf{A})\leq\frac{\mathrm{tr}(\mathbf{A})+\sqrt{n(n-1)\|\mathbf{A}\|_F^2-(n-1)\mathrm{tr}(\mathbf{A})^2}}{n},\\
&\lambda(\mathbf{A})\geq\frac{n}{\mathrm{tr}(\mathbf{B})+\sqrt{n(n-1)\|\mathbf{B}\|_F^2-(n-1)\mathrm{tr}(\mathbf{B})^2}}.
\end{aligned}
\end{equation}
where $\kappa(\cdot)$ denotes the maximum eigenvalue of the matrix, and $\mathbf{B}=\mathbf{A}^{-1}$.
\end{theorem}

\begin{Proof}
Let us assume that $\mathrm{tr}(\mathbf{A})$ and $\|\mathbf{A}\|_F^2$ are known, then we consider the following problem
\begin{equation}
\begin{aligned}
&~~~~~~~~~~~~~~~~~~~~~~\bm{\eta}^*=\arg\max_{\bm{\eta}\in \mathcal{S}}\bm{\eta},\\
&\mathcal{S}=\left\{\bm{\eta}\in\mathbb{R}^n|\bm{1}^\mathrm{T}\bm{\eta}=\mathrm{tr}(\mathbf{A}),\bm{\eta}^\mathrm{T}\bm{\eta}=\|\mathbf{A}\|_F^2,\bm{\eta}\geq\bm{0}\right\}.
\end{aligned}
\end{equation}

According to the Karush-Kuhn-Tucker conditions, we have
\begin{align}
\!\!\!\!\kappa(\mathbf{A})\leq\eta ^*=\frac{\mathrm{tr}(\mathbf{A})\!+\!\sqrt{n(n\!-\!1)\|\mathbf{A}\|_F^2\!-\!(n\!-\!1)\mathrm{tr}(\mathbf{A})^2}}{n},
\end{align}

Similarly, apply the bound to $\mathbf{B}=\mathbf{A}^{-1}$, we have
\begin{align}
\!\!\!\!\!\lambda(\mathbf{A})\!=\!\frac{1}{\kappa(\mathbf{B})}\!\geq\!\frac{n}{\mathrm{tr}(\mathbf{B})\!+\!\!\sqrt{n(n\!-\!1)\|\mathbf{B}\|_F^2\!-\!(n\!-\!1)\mathrm{tr}(\mathbf{B})^2}}.\!\!
\end{align}

The Proof is finished.
\end{Proof}
\section{Numerical Results}

In this section, we present the numerical simulation results by comparing the performance in terms of mutual information. The number of the receiving antennas is set as $N_R=4$. The elements of the channel matrices corresponding to the desired and interfering signals are configured to follow Gaussian distribution, i.e., $\mathbf{H}_D(i,j)\in \mathcal{CN}(0,\sigma^2_D),\mathbf{H}_I(i,j)\in \mathcal{CN}(0,\sigma^2_I),\forall i \in N_R, j\in N_D,N_I$. The AWGN is assumed to have unit power, then signal to noise ratio (SNR) and interference to noise ratio (INR) are defined as
\begin{align}
SNR=\sigma^2_D,~~~~~INR=\sigma^2_I.
\end{align}
The desired signal is set to be a single stream transmission $N_D=1$, and $Q$-QAM signals are transmitted. The linear minimum mean square detector is performed the calculate the log likelihood ratio (LLR). The mutual information between the bits and their corresponding LLRs which are defined as
\begin{align}
MI=\log_2 Q-\sum_{i=1}^{\log_2 Q}\mathbb{E}_{b_i,L_i}\log_2\left[1+\mathrm{e}^{-(1-2b_i)L_i}\right],
\end{align}
is utilized to evaluate the performance, where $b_i$ and $L_i$ denote the $i$-th bit and its corresponding LLR, respectively.

\begin{figure}
\centering
\includegraphics[width=3.4in]{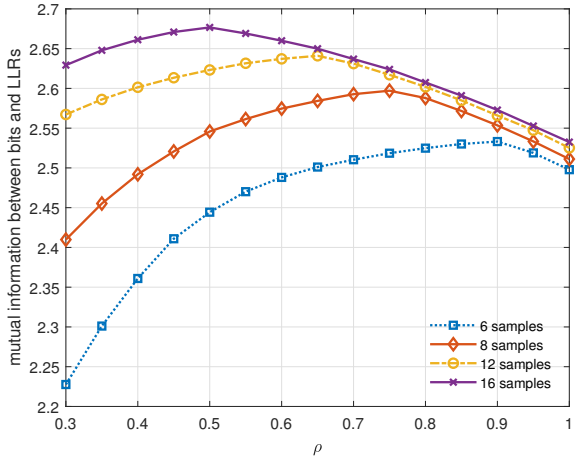}
\caption{Mutual information between bits and LLRs with respect to $\rho$, for 4 receiving antennas, one stream desired signal with 16QAM and SNR = 10 dB, one stream interfering signal with INR = 20 dB; the number of samples used for the covariance matrix estimation are set as 6, 8, 12, 16, respectively.}
\label{fig:1}
\end{figure}

The mutual information between bits and LLRs with different $\rho$'s are shown in Fig. \ref{fig:1}, where the desired signal is set as 16QAM and SNR = 10 dB, and the interfering signal is set to be one single stream transmission with INR = 20 dB. The number of the samples that are used for estimating the covariance matrix is set as 6, 8, 12 and 16, respectively. It can be seen that there exists a best $\rho$ that achieves the highest mutual information, and the best $\rho$ decreases with the increase of the number of sample.

\begin{figure}[t]
\centering
\includegraphics[width=3.4in]{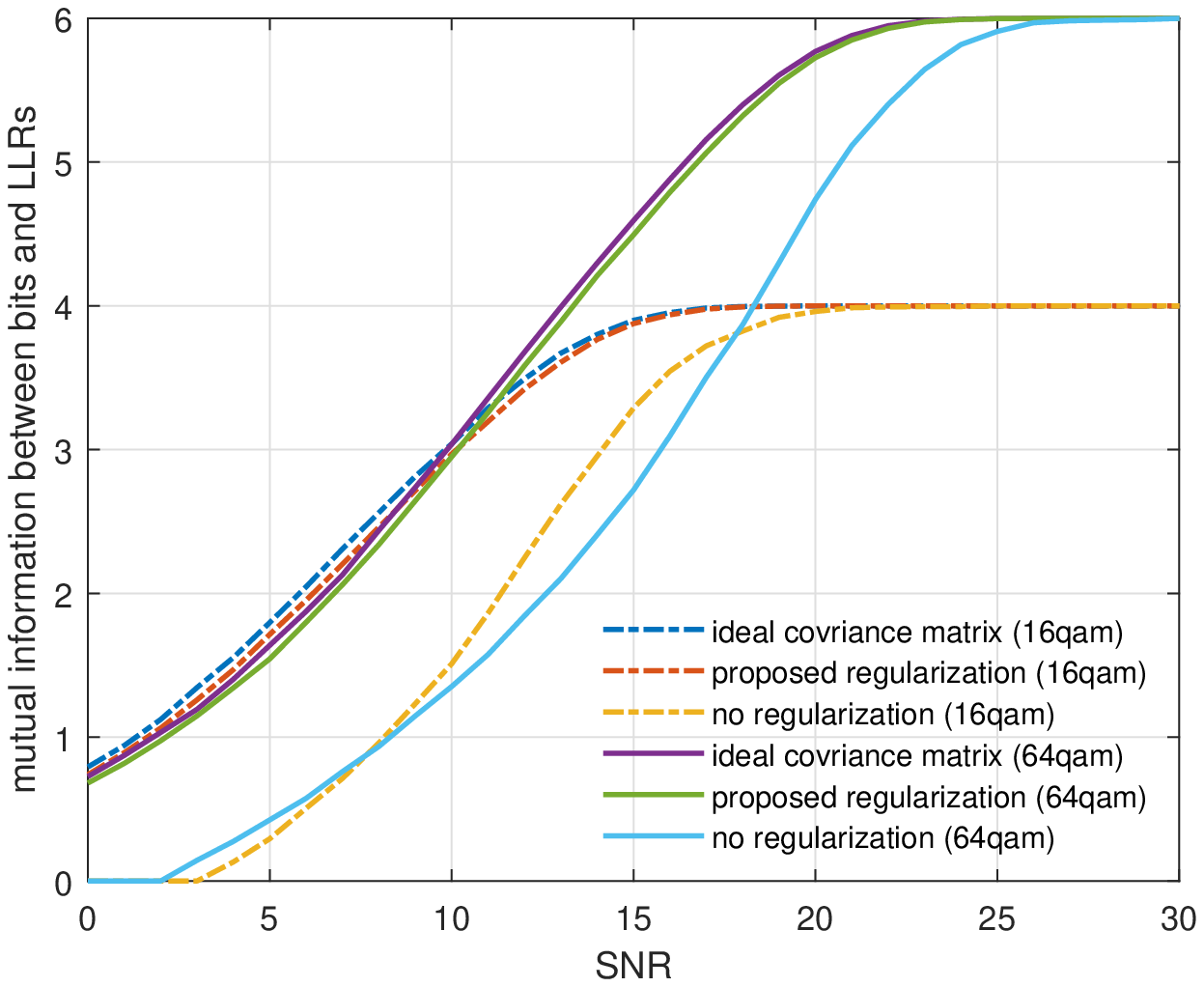}
\caption{Mutual information between bits and LLRs with respect to SNR, for 4 receiving antennas, one stream desired signal with 16QAM and 64QAM, respectively, and one stream interfering signal with INR = 0 dB; the number of samples used for the covariance matrix estimation is set to be 8.}
\label{fig:2}
\end{figure}

In Fig. \ref{fig:2}, we show the mutual information between bits and LLRs with respect to SNR, where the desired signal is set as 16QAM and 64QAM, and the interfering signal is set to be one single stream transmission with INR = 0 dB. The number of the samples that are used for estimating the covariance matrix is set to be 8. The performance of the proposed estimator has been compared against two competitors: the oracle estimator with the true covariance matrix and the estimator with no regularization. The $\rho^*$ in our proposed regularization is obtained from (\ref{eq:rho*}). The proposed regularization exhibits a much more significant gain over the estimator with no regularization, and is very close to the oracle in terms of mutual information.

\begin{figure}[t]
\centering
\includegraphics[width=3.4in]{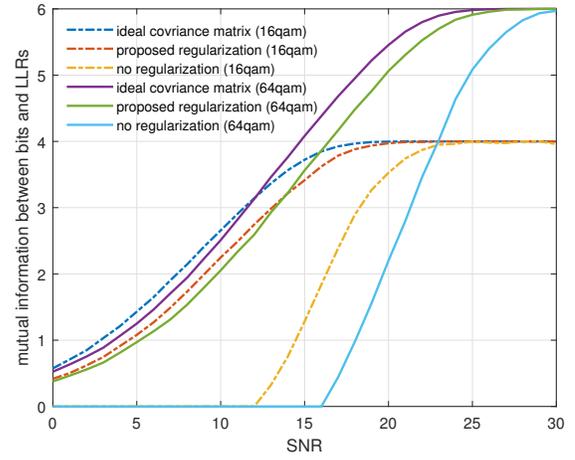}
\caption{Mutual information between bits and LLRs with respect to SNR, for 4 receiving antennas, one stream desired signal with 16QAM and 64QAM, respectively, and two streams interfering signal with INR = 20 dB; the number of samples used for the covariance matrix estimation is set to be 6.}
\label{fig:3}
\end{figure}

In this case, the interfering signal is set to be two streams with INR = 20 dB, and the number of the samples that are used for estimating the covariance matrix is set to be 6. Fig. \ref{fig:3} reports the mutual information between bits and LLRs with respect to SNR, where the desired signal is set as 16QAM and 64QAM. The oracle estimator with the true covariance matrix and the estimator with no regularization are also simulated for comparison. The proposed approach could be a means to improve regularization towards optimality in the mutual information sense, which shows that the minimum eigenvalue is an effective criterion.

\section{Conclusion}
We have considered the topic of IRC receiver which requires the knowledge of the covariance matrix including the interference signals and noise. In order to solve the regularized SCM estimation problem with limited samples, a minimum eigenvalue based regularization parameter optimization is developed. Different from traditional methods that aim at minimizing the MSE, the proposed method goes straight at the objective of optimizing the final performance of an IRC receiver. Simulation results show demonstrate that the proposed approach is effective and can approach the performance of the ideal oracle estimator in terms of the mutual information metric.

\end{document}